\documentclass[
twocolumn, preprintnumbers, showpacs,
amssymb,amsmath,aps,floatfix,prl,footnoteinbib]{revtex4-1}
  \usepackage{latexsym,bm,amsmath,amssymb,amsfonts}
\newcommand{\beq}{\begin{eqnarray}}
\newcommand{\eeq}{\end{eqnarray}}

\newcommand{\nn}{\nonumber \\}

\usepackage{graphicx}
\usepackage{amsmath,amsthm,amssymb}

\begin{document}

\preprint{YITP-14-6}

\title{Exact analytical solutions of second-order conformal hydrodynamics}
\author{Yoshitaka Hatta$^1$}
\author{Jorge Noronha$^2$}
\author{ Bo-Wen Xiao$^3$}

\affiliation{\vspace{3mm}
$^1$Yukawa Institute for Theoretical Physics, Kyoto University, Kyoto 606-8502, Japan \\ $^2$Instituto de F\'isica, Universidade de S\~ao Paulo, C.P. 66318, 05315-970 S\~ao Paulo, SP, Brazil \\ $^3$Key Laboratory of Quark and Lepton Physics (MOE) and Institute
of Particle Physics, Central China Normal University, Wuhan 430079, China}

\date{\today}
\vspace{0.5in}
\begin{abstract}
We present some exact solutions of relativistic second-order hydrodynamic equations in theories with conformal symmetry. Starting from a spherically expanding solution in ideal hydrodynamics, we take into account general conformal second-order corrections, and construct, for the first time, fully analytical axisymmetric exact solutions including the case with nonzero vorticity. These solutions are time-reversible despite having a nonvanishing shear stress tensor and provide a useful quantitative measure of the second-order effects in relativistic hydrodynamics.
\end{abstract}
\maketitle

Given the apparent success of the hydrodynamic description of  the quark-gluon plasma formed in ultrarelativistic heavy-ion collisions at RHIC and the LHC \cite{Heinz:2013th}, significant progress has been achieved in the foundation and applications of relativistic hydrodynamics. In particular, there have been many attempts \cite{DeGroot:1980dk,Baier:2007ix,Bhattacharyya:2008jc,Koide:2006ef,PeraltaRamos:2009kg,Denicol:2011fa,Denicol:2012cn,Tsumura:2012kp} to derive a consistent theory of second-order relativistic hydrodynamics which generalizes the original Israel-Stewart theory \cite{Israel:1979wp}. Some of them have already been implemented in numerical codes for practical applications in heavy-ion collisions \cite{Luzum:2008cw}.

Second-order hydrodynamic equations typically contain many new variables compared with ideal hydrodynamics, and it seems an impossible task to solve them analytically. Indeed, although there are a number of exact solutions of relativistic ideal hydrodynamics known in the literature (see, e.g., \cite{Bjorken:1982qr,Biro:2000nj,Csorgo:2003rt,Csorgo:2006ax,Bialas:2007iu,Beuf:2008vd,
 Fouxon:2008ik,Nagy:2009eq,Liao:2009zg}), with few exceptions \cite{Gubser:2010ze,Gubser:2010ui,Marrochio:2013wla} there has been little hope of generalizing them to include even the first-order (Navier-Stokes) corrections in the relativistic domain, let alone second-order ones.

In this Letter, we  expand the current knowledge of analytic solutions in relativistic hydrodynamics by presenting the first nontrivial exact solutions to the general second-order conformal hydrodynamic equations including the case with nonzero vorticity.
 Our solutions are  explicit, have a surprisingly simple mathematical structure, and are valid for rather generic values of the transport coefficients involved. They thus serve as a useful reference point of the future study of various second-order effects in relativistic hydrodynamics.

Our strategy to construct solutions of conformal hydrodynamics is similar to that of Gubser \emph{et al}. \cite{Gubser:2010ze,Gubser:2010ui,Marrochio:2013wla}.
We start with the following rewriting of the  Minkowski metric ($x_\perp\equiv \sqrt{x^2+y^2}$)
\beq
ds^2 &=& -dt^2 + dz^2 + dx_\perp^2 + x_\perp^2 d\phi^2 \nn
 &=&x_\perp^2 \left( \frac{-dt^2 + dz^2 + dx_\perp^2}{x_\perp^2} + d\phi^2 \right)\,. \label{conformal}
\eeq
This shows that the Minkowski space is conformal to $AdS_3 \times S^1$ up to a  Weyl rescaling factor $x_\perp^2$. We shall solve the hydrodynamic equations in the latter space, and then conformally map the solution to the Minkowski space. The solution that we are after is simply described in the so-called global coordinates of $AdS_3$ \cite{footnote1} where
 the metric $d\hat{s}^2 \equiv ds^2/x_\perp^2$ takes the form
\begin{equation}
d\hat{s}^2 \!= -\cosh^2 \rho \, d\tau^2 + d\rho^2 + \sinh^2 \rho d\Theta^2 + d\phi^2. \label{met}
\end{equation}
 $\rho$ is defined in terms of Minkowski coordinates $(t,r,x_\perp)$ via
\beq
\cosh\rho = \frac{1}{2Lx_\perp} \sqrt{(L^2+(r+t)^2)(L^2+(r-t)^2)}\,, \nonumber
\eeq
 where $r=\sqrt{z^2+x_\perp^2}$ and $L$ is the radius parameter of $AdS_3$.
 We shall denote variables  with a ``hat" for quantities in  global coordinates $\hat{x}^\mu=(\tau,\rho,\Theta,\phi)$.
In this coordinate system, we consider hydrostatic fluid static in `time' $\tau$. This is  characterized by the flow velocity ($\hat{u}^\mu \hat{u}_\mu=-1$)
\beq
\hat{u}_\tau=-\cosh\rho\,, \qquad   \hat{u}^\rho=\hat{u}^\Theta=\hat{u}^\phi=0\,.
\label{hat}
\eeq
The corresponding flow velocity in Minkowski coordinates is obtained by $u_\mu=-x_\perp \frac{d\hat{x}^\nu}{dx^\mu}\hat{u}_\nu$ ($x_\perp$ is the Weyl rescaling factor \cite{Gubser:2010ze}) and reads
\beq
u_t &=& -\frac{L^2+r^2+t^2}{\sqrt{(L^2+(r+t)^2)(L^2+(r-t)^2)}}\,,  \nn
\vec{u}&=& \frac{2t\vec{r}}{\sqrt{(L^2+(r+t)^2)(L^2+(r-t)^2)}}\,. \label{ut}
\eeq
This is a radially expanding spherically symmetric flow with vanishing shear tensor
$\sigma^{\mu\nu}\equiv \Delta^{\mu\nu\alpha\beta}\nabla_\alpha u_\beta= 0$
 where  $\Delta^{\mu\nu\alpha\beta}\!\equiv \!\frac{1}{2}(\Delta^{\mu\alpha}\Delta^{\nu\beta}
+\Delta^{\mu\beta}\Delta^{\nu\alpha})-\frac{1}{3}\Delta^{\mu\nu}\Delta^{\alpha\beta} $, $\Delta^{\mu\nu}\equiv g^{\mu\nu}+u^\mu u^\nu$, and nonvanishing expansion rate
\beq
\theta\equiv \nabla_\mu u^\mu = 3\frac{u^r}{r}\,.
\label{exp}
\eeq
(Note that $\hat{\theta}=0$ since the flow is static in global coordinates. This quantity does not transform homogeneously under the Weyl rescaling.)

With this flow velocity $\hat{u}^\mu$, the energy momentum tensor of
viscous conformal fluids (i.e., relativistic fluids in which the energy density $\hat{\epsilon}$ is given by $\hat{\epsilon} =3\hat{p}$ with $\hat{p}$ being the pressure) is written as $\hat{T}^{\mu\nu}=\hat{\epsilon} \,\hat{u}^\mu \hat{u}^\nu+\hat{\epsilon} \hat{\Delta}^{\mu\nu}/3+\hat{\pi}^{\mu\nu}$. $\hat{\pi}^{\mu\nu}$ is the shear stress tensor that is symmetric and traceless, and is typically chosen to be orthogonal to the flow  $\hat{u}_\mu \hat{\pi}^{\mu\nu}=0$ (the so-called Landau frame \cite{landau}). It enters the energy-momentum conservation equations $\hat{\nabla}_\mu \hat{T}^{\mu\nu}=0$ (in $\hat{x}$-coordinates) as follows
\beq
\hat{D}\hat{\epsilon}
&=& 0\,,
\label{energyconserv} \\
4\hat{\epsilon}\, \hat{D}\hat{u}^\mu +\hat{\Delta}^{\mu\nu}\hat{\nabla}_\nu\epsilon+3 \hat{\Delta}^{\mu}_\nu \hat{\nabla}_\alpha \hat{\pi}^{\nu\alpha}&=& 0\,,
\label{vectorconserv}
\eeq
where the co-moving derivative is $\hat{D}\equiv \hat{u}_\mu \hat{\nabla}^\mu$ and we have already set $\hat{\theta}=\hat{\sigma}^{\mu\nu}=0$.
 (\ref{energyconserv}) and (\ref{vectorconserv}) should be supplemented with the constitutive  equation for $\hat{\pi}_{\mu\nu}$. Its most general form is rather complicated \cite{Denicol:2012cn}, but when $\hat{\sigma}^{\mu\nu}=0$ it can be written as \cite{Baier:2007ix,Bhattacharyya:2008jc}
\beq
\hat{\pi}^{\mu\nu} &=& -\frac{\tau_\pi}{\hat{\epsilon}^{1/4}} \hat{\Delta}^\mu_\alpha \hat{\Delta}^\nu_\beta \hat{D} \hat{\pi}^{\alpha\beta}  + \frac{\lambda_1}{\hat{\epsilon}}\hat{\pi}^{\langle \mu}_{\ \ \lambda}\hat{\pi}^{\nu\rangle\lambda} \nn
 && \quad \ +\frac{\lambda_2}{\hat{\epsilon}^{1/4}}\hat{\pi}^{\langle \mu}_{\ \ \lambda}\hat{\Omega}^{\nu\rangle\lambda}+ \lambda_3\hat{\epsilon}^{1/2}\hat{\Omega}^{\langle \mu}_{\ \  \lambda}\hat{\Omega}^{\nu\rangle \lambda}\nn
&&+\kappa \hat{\epsilon}^{1/2}\left(\hat{\mathcal{R}}^{\langle \mu\nu\rangle} -2\hat{u}_\alpha \hat{\mathcal{R}}^{\alpha\langle\mu\nu\rangle\beta}\hat{u}_\beta \right)\,,
\label{c}
\label{definepi}
\eeq
where $\hat{\mathcal{R}}^{\mu\nu\alpha\beta}$ is the Riemann curvature tensor, $A^{\langle \mu\nu\rangle} \equiv \Delta^{\mu\nu\alpha\beta}A_{\alpha\beta}$, 
 and $\hat{\Omega}^{\mu\nu}\equiv \frac{1}{2}\hat{\Delta}^{\mu\alpha}\hat{\Delta}^{\nu\beta}(\hat{\nabla}_\alpha \hat{u}_\beta -\hat{\nabla}_\beta \hat{u}_\alpha)$ is the vorticity tensor.
 The transport coefficients $\tau_\pi$, $\kappa$, $\lambda_i$ ($i=1,2,3$) are  dimensionless, and are rescaled by the appropriate power of $\hat{\epsilon}$. This is because these coefficients are dimensionful in the original Minkowski space and are proportional to  $\epsilon$ to some power in a conformal fluid. After the Weyl rescaling, this is converted to a power of $\hat{\epsilon}=x_\perp^4 \epsilon$.

For our purposes, it is important to emphasize that $\hat{\pi}^{\mu\nu}$ are treated as \emph{independent} variables which should be determined self-consistently by the constitutive equation (\ref{c}). This follows the spirit of the original Israel-Stewart approach, and has been recently put on a firm ground in \cite{Denicol:2012cn} where (\ref{c}) was derived via the consistent truncation of the Boltzmann equation doubly expanded in powers of $\sigma^{\mu\nu}$ and $\pi^{\mu\nu}$. This is indeed crucial for our problem since the commonly employed lowest-order substitution $\hat{\pi}^{\mu\nu} \leftrightarrow -2\eta \hat{\sigma}^{\mu\nu}$ ($\eta$ is the shear viscosity) fails in this case since $\hat{\sigma}^{\mu\nu}=0$.

Since the background space-time is conformally equivalent to flat space, the term proportional to $\kappa$ vanishes identically. The vorticity tensor also vanishes $\hat{\Omega}^{\mu\nu}=0$  for the flow velocity (\ref{hat}). Moreover, (\ref{energyconserv}) shows that $\hat{\epsilon}$ (hence also $\hat{\pi}^{\mu\nu}$) does not depend on  $\tau$. The equation for $\hat{\pi}^{\mu\nu}$ then simplifies to
 \beq
 \hat{\pi}^{\mu\nu} = \frac{\lambda_1}{\hat{\epsilon}} \hat{\pi}^{\langle \mu}_{\ \ \lambda}\hat{\pi}^{\nu\rangle\lambda} \,. \label{nont}
 \eeq
 Assuming $\hat{\pi}^{\mu\nu}$ is diagonal, we find the solution
 \beq
 (\hat{\pi}^{\rho\rho},\sinh^2\rho\, \hat{\pi}^{\Theta\Theta}, \hat{\pi}^{\phi\phi})\!=\!\frac{\hat{\epsilon}}{\lambda_1}\!\times\! \begin{cases}
  (-1,-1,2)\,, \\
 (-1,2,-1)\,, \\
   (2,-1,-1)\,.
   \end{cases} \label{case}
 \eeq
 Plugging (\ref{case}) into (\ref{vectorconserv}), we see that
the $\hat{x}^\mu=\tau$ component is trivially satisfied and the $\hat{x}^\mu=\Theta, \phi$ components reduce to $
 \partial_\Theta \hat{\epsilon} =\partial_\phi \hat{\epsilon}=0$.
The $\hat{x}^\mu=\rho$ component is nontrivial and reads
\beq
&&\partial_\rho \hat{\epsilon} + 4\hat{\epsilon}\tanh\rho +3\Bigl( \partial_\rho \hat{\pi}^{\rho\rho}+ \left(\tanh\rho+\coth \rho\right)\hat{\pi}^{\rho\rho}
\nn && \qquad \qquad \qquad -\sinh\rho\cosh\rho \hat{\pi}^{\Theta\Theta} \Bigr)=0\,. \label{rho}
\eeq
This can be easily solved as
  \beq
\hat{\epsilon}\propto \begin{cases}
\left(\frac{1}{\cosh^2\rho}\right)^{2+\frac{9}{2(\lambda_1-3)}}\,, \\
\left(\frac{1}{\cosh^2\rho}\right)^2 (\tanh^2\rho)^{\frac{9}{2(\lambda_1-3)}}\,, \\
\left(\frac{1}{\cosh^2\rho}\right)^2 \left(\frac{\tanh^2\rho}{\cosh^2\rho}\right)^{-\frac{9}{2(\lambda_1+6)}}\,. \end{cases}
\label{sol}
 \eeq
The corresponding energy density in Minkowski space reads
\begin{widetext}
 \beq
  \epsilon\propto \begin{cases}
  \frac{1}{(L^2+(t+r)^2)^2(L^2+(t-r)^2)^2}
  \left(\frac{4L^2x_\perp^2}{(L^2+(t+r)^2)(L^2+(t-r)^2)}\right)^{\frac{9}{2(\lambda_1-3)}}\,,
  \\
\frac{1}{(L^2+(t+r)^2)^2(L^2+(t-r)^2)^2}
\left(1-\frac{4L^2x_\perp^2}{(L^2+(t+r)^2)(L^2+(t-r)^2)}\right)^{\frac{9}{2(\lambda_1-3)}}\,, \\
\frac{1}{(L^2+(t+r)^2)^2(L^2+(t-r)^2)^2}\left( \frac{4L^2x_\perp^2\bigl((L^2+(t+r)^2)(L^2+(t-r)^2)-4L^2x_\perp^2\bigr)}{(L^2+(t+r)^2)^2(L^2+(t-r)^2)^2}
\right)^{-\frac{9}{2(\lambda_1+6)}}\,. \label{viscous}
\end{cases}
 \eeq
\end{widetext}
The corresponding temperature  is given by $T(t,r)\propto \epsilon^{1/4}(t,r)$.

As a consistency check, we have numerically confirmed that $\int T^{00}d^3\vec{r}$ is constant in $t$ for all these three solutions. For the first two solutions we need to require $\lambda_1>3$ in order for the total energy to be finite.

Eq.\ (\ref{case}) shows that $\lambda_1$ essentially plays the role of the Reynolds number $Re^{-1}=\sqrt{\hat{\pi}^{\mu\nu}\hat\pi_{\mu\nu}}/\hat{\epsilon} \sim 1/\lambda_1$. Since the constitutive equation (3) involves the expansion in inverse powers of the Reynolds number \cite{Denicol:2012cn}, consistency requires that $\lambda_1$ has to be large. In particular, the ideal hydro limit corresponds to $\lambda_1\to \infty$, contrary to the naive expectation $\lambda_1\to 0$. Indeed, in the limit $\lambda_1\to \infty$ the above solutions reduce to the spherically expanding solution in ideal hydrodynamics previously obtained by Nagy using a different method \cite{Nagy:2009eq}. [The particular solution with $L=0$ was found earlier \cite{Csorgo:2006ax}.] We see that the finite-$\lambda_1$ corrections break rotational symmetry down to axial symmetry ($\phi$-rotation).
 In Fig.~1, we plot the time-evolution of the energy density profile $\epsilon(t,x_\perp, z=0)$ for the three solutions.
\begin{figure}[htbp]
  \begin{center}
   \includegraphics[width=65mm]{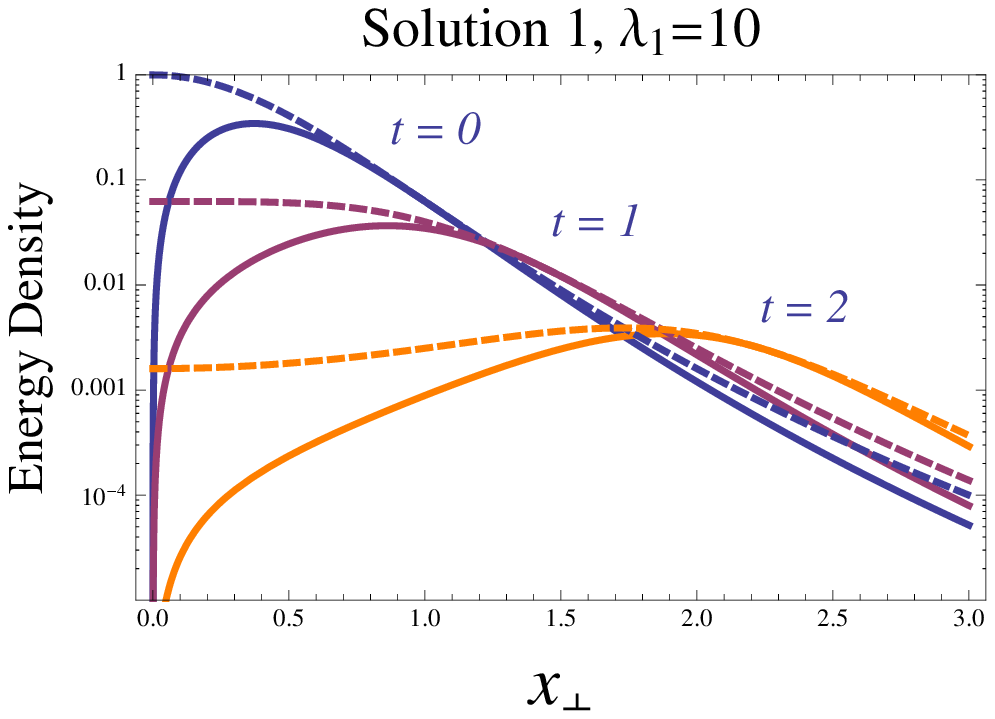}
  \end{center}
\begin{center}
   \includegraphics[width=65mm]{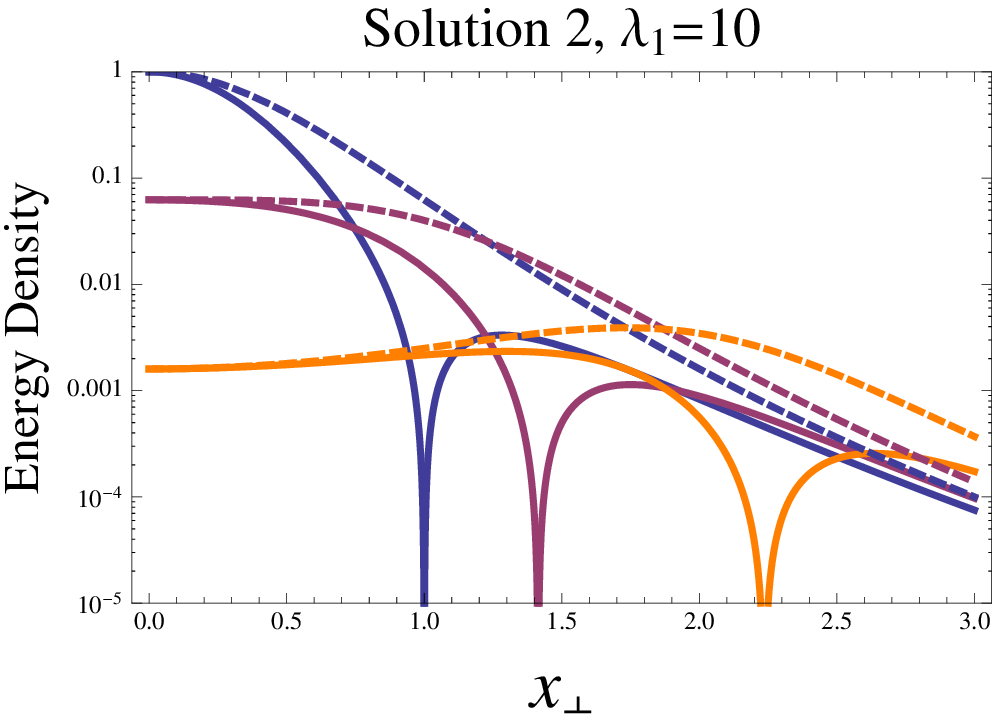}
   \end{center}
  \begin{center}
   \includegraphics[width=65mm]{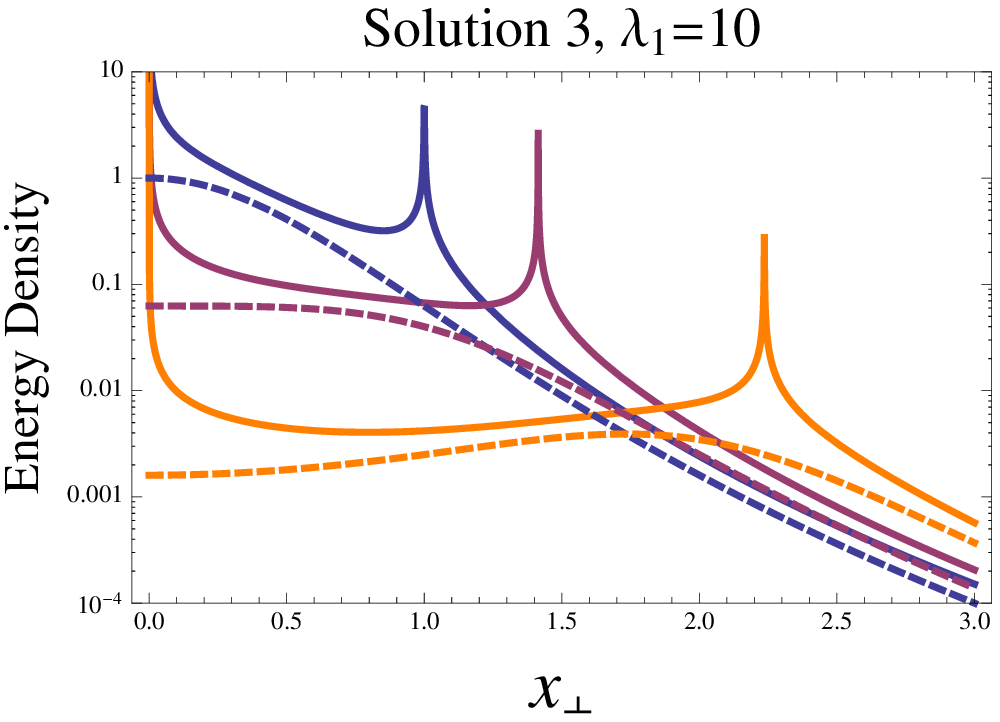}
  \end{center}
 \caption{The time-evolution of the energy density profile $\epsilon(x_\perp)$ at $z=0$ and $t=0,1,2$  for the three solutions in (\ref{viscous}). We set $L=1$ and $\lambda_1=10$. The dashed lines represent the ideal solution in the limit $\lambda_1\to \infty$.}
\end{figure}

Note that the solutions (\ref{viscous}) are invariant under $t\to -t$, that is, they are \emph{time-reversible} despite $\pi^{\mu\nu}\neq 0$. A simple look at the energy-momentum conservation equations tells us that time reversal invariance is broken if $\pi_{\mu\nu}$ is not even under this operation. In fact, when $t\to -t$ we have that the spatial flow velocity changes as $\vec{u} \to -\vec{u}$, while $\theta \to -\theta$, $\sigma_{\mu\nu}\to -\sigma_{\mu\nu}$. Thus, in the Navier-Stokes approximation $\pi_{\mu\nu} \sim -2\eta \sigma_{\mu\nu}$, one can clearly see that time reversal invariance is broken, which is of course associated with the production of entropy \cite{landau}. However, our solutions are static in $\tau$ and dissipationless $\hat{\sigma}^{\mu\nu}=\hat{\theta}=0$. Clearly, then, a nontrivial solution of (\ref{nont}) implies that $\hat{\pi}^{\mu\nu}$ is even under time-reversal. Thus, it is indeed possible to find time-reversible fluid configurations in second-order hydrodynamics.
Presumably, in the context of kinetic theory, our solutions may correspond to some kind of a nontrivial fixed point of the Boltzmann equation that never reaches local thermal equilibrium. This certainly deserves further study.\\

In \cite{Nagy:2009eq}, Nagy also derived an exact ideal-fluid solution with rotation. It is easy to accommodate this solution in our framework. For  fluid rotating around the $z$-axis, we just turn on the $\phi$-component of the velocity.
\begin{equation}
\hat{u}_\tau= \!\frac{-\cosh^2\rho}{\sqrt{\cosh^2\rho -\omega^2}}\,, \
\hat{u}_\phi =\!\frac{\omega}{\sqrt{\cosh^2\rho -\omega^2}}\,. \label{newu}
\end{equation}
 with the obvious constraint $1\ge \omega \ge 0$.
The corresponding flow velocity in Minkowski space is
\beq
u_t&=&-\frac{L^2+r^2+t^2}{\sqrt{(L^2+t^2+r^2)^2-4r^2t^2 -4\omega^2L^2x_\perp^2}}\,, \nn
\vec{u}&=&\frac{2t\vec{r} + 2\omega L (\vec{r}\times \vec{e}_z)}{\sqrt{(L^2+t^2+r^2)^2-4r^2t^2-4\omega^2L^2x_\perp^2}}\,,
\eeq
in agreement with \cite{Nagy:2009eq}. The modified flow velocity (\ref{newu}) still satisfies $\hat{\theta}=\hat{\sigma}^{\mu\nu}=0$.
The energy density is given by
\beq
\hat{\epsilon}\propto \frac{1}{(\cosh^2\rho -\omega^2)^2}\,, \label{newhat}
\eeq
or in Minkowski space,
\begin{equation}
\epsilon \propto\!
\frac{1}{((L^2+t^2+r^2)^2-4r^2t^2 -4\omega^2L^2 x_\perp^2)^2}\,.
\end{equation}
Naturally, this solution has nonzero vorticity
\begin{equation}
\hat{\Omega}_{\tau\rho}=\!\frac{-\omega^2\cosh\rho\sinh\rho}{(\cosh^2\rho -\omega^2)^{3/2}}\,, \ \   \hat{\Omega}_{\phi\rho}\!=\!\frac{\omega \cosh\rho \sinh\rho}{(\cosh^2\rho -\omega^2)^{3/2}}\,. \nonumber
\end{equation}
We now include second-order corrections to this solution keeping the flow velocity (\ref{newu}) unchanged.
 Temporarily assuming $\tau_\pi=\lambda_2=0$, we find that the solution to (\ref{c})
 \beq
\hat{\pi}^{\mu\nu}&=& \frac{\lambda_1}{\hat{\epsilon}} \hat{\pi}^{\langle \mu}_{\ \ \lambda}\hat{\pi}^{\nu\rangle\lambda} +
\lambda_3\sqrt{\hat{\epsilon}}\, \hat{\Omega}^{\langle \mu}_{\ \  \lambda}\hat{\Omega}^{\nu\rangle \lambda}\,,
 \eeq
 is given by
 \beq
&& (\hat{\pi}^{\rho\rho}, \sinh^2\rho\hat{\pi}^{\Theta\Theta}, \hat{\pi}^{\phi\phi})\!=\!\frac{\hat{\epsilon}}{\lambda_1}\left(\alpha, \beta, \frac{\gamma\cosh^2\rho}{ \cosh^2\rho-\omega^2}\right)\,,  \nn
&& \qquad \hat{\pi}^{\tau\tau}= \frac{\omega}{\cosh^2\rho} \hat{\pi}^{\tau\phi}=\frac{\omega^2}{\cosh^4\rho} \hat{\pi}^{\phi\phi}\,. \label{den}
 \eeq
The parameters $(\alpha,\beta,\gamma)$ in (\ref{den}) can be any of the following four possibilities
\beq
&&(\alpha,\beta,\gamma) \!=\!\begin{cases}
\left(\frac{1-\sqrt{9-4f}}{2}, -1, \frac{1+\sqrt{9-4f}}{2}\right)\,,\\
\left(\frac{1+\sqrt{9-4f}}{2},-1,\frac{1-\sqrt{9-4f}}{2} \right)\,,
\end{cases} \label{ca1}\\
&&\alpha=\gamma=-\frac{\beta}{2}=\begin{cases}
\frac{1}{2}\left(-1-\sqrt{1+4f/3}\right)\,, \\
\frac{1}{2}\left(-1+\sqrt{1+4f/3}\right)\,,
\end{cases}
 \label{ca2}
\eeq
where we defined
\beq
f\equiv \frac{\lambda_1\lambda_3\omega^2\sinh^2\rho}{\sqrt{\hat{\epsilon}}(\cosh^2\rho-\omega^2)^2}\,.
\label{ff}
\eeq
In the $f\to 0$ limit, the first three solutions in (\ref{ca1})-(\ref{ca2}) reduce to (\ref{case}), while the last solution
reduces to the one with $\lambda_1=0$.
 Below we consider only the last two solutions (\ref{ca2}) because they satisfy $ \hat{\Delta}^\mu_\alpha \hat{\Delta}^\nu_\beta \hat{D} \hat{\pi}^{\alpha\beta}=\hat{\pi}^{\langle \mu}_{\ \ \lambda}\hat{\Omega}^{\nu\rangle\lambda}=0$, namely, they are solutions even when $\tau_\pi, \, \lambda_2\neq 0$. 

Substituting (\ref{den}) and (\ref{ca2}) into (\ref{vectorconserv}), we are left with the following nonlinear
differential equation
\beq
&&\partial_\rho \hat{\epsilon}+\!\frac{4 \cosh \rho \sinh \rho}{\cosh^2 \rho -\omega^2}\hat{\epsilon}\!+\! 3\!\left[ \partial_\rho \hat{\pi}^{\rho\rho}\!+\!\frac{4 \cosh \rho \sinh \rho}{\cosh^2 \rho -\omega^2}\hat{\pi}^{\rho\rho} \right] \nn && \qquad \qquad+\frac{9 (1-\omega^2)\coth \rho}{\cosh^2 \rho -\omega^2}\hat{\pi}^{\rho\rho}=0\,. \label{non}
\eeq
To solve this, we employ an ansatz
 \begin{equation}
\hat{\epsilon}=\frac{A^2(\rho)\sinh^4\rho}{(\cosh^2\rho -\omega^2)^4}\,,
\label{con}
\end{equation}
 with which we can write
 $\hat{\pi}^{\rho\rho}=b(\rho)\hat{\epsilon}$ where $b(\rho)$ is the root of
 \begin{equation}
A(\rho)=\frac{\lambda_3 \omega^2}{3b(\rho)(\lambda_1 b(\rho)+1)}\,. \label{b2}
\end{equation}
 First we find the special solution where $A$ is a constant. In this case, (\ref{non}) requires either $\omega=1$ or $b=-\frac{4}{21}$.
 When $\omega=1$, the solution turns out to be the same as the ideal one
 \beq
 \hat{\epsilon}= \frac{A^2}{\sinh^4\rho}\,,
 \eeq
  where $A$ is arbitrary.
 On the other hand, when  $b= -\frac{4}{21}$,
 \beq
 A=\frac{7\lambda_3\omega^2}{4\left(\frac{4}{21}\lambda_1-1\right)}\,, \label{spe}
 \eeq
 is the solution. 
 Note that if we take the limits $\omega\to 0$ and $\lambda_1\to \frac{21}{4}$ in (\ref{spe}) such that $A$ remains finite,  (\ref{con})  reduces to the second solution in (\ref{sol}).

 To find general solutions, we now take into account the $\rho$-dependence of $A(\rho)$.
 (\ref{non}) can be cast into a differential equation for $b(\rho)$
\begin{equation}
\frac{9\lambda_1 b^2+4\lambda_1 b+3b+2}{b(\lambda_1 b+1)(4+21b)}\partial_\rho b=\frac{(1-\omega^2)\coth \rho}{\cosh^2\rho -\omega^2}\,.
\end{equation}
The above equation can be integrated as
\begin{equation}
b\left(21b+4\right)^{e_1}\left(\lambda_1b+1\right)^{e_2}=\textrm{C} \frac{\sinh^2\rho}{\cosh^2\rho -\omega^2}\, , \label{gs}
\end{equation}
where $e_1=\frac{105-32\lambda_1}{7(4\lambda_1-21)}$,  $e_2=1+\frac{9}{4\lambda_1-21}$ and $\textrm{C}$ is the integration constant.  Given $b^{\ast} (\rho)$ as the solution to (\ref{gs}), we can obtain the energy density accordingly.
 In the limit $\lambda_1\to \infty$, (\ref{gs}) indicates that $b\to 0$, and we  recover the ideal solution  (\ref{newhat}).\\

In conclusion,  we have found several novel exact  solutions to the  second-order conformal hydrodynamic equations in which $\pi^{\mu\nu}$ is treated as independent variable. The ideal-fluid limit of these solutions reduces to previously known results. Our solutions encode very interesting nonlinear effect as well as the vorticity contribution, which has never been studied before. We showed that the second-order equations allow for nontrivial time-reversible solutions in which $\pi^{\mu\nu}$ cannot be approximated by its Navier-Stokes limit.
Although these solutions are rather special, they cannot be ruled out as nonphysical simply using the second law of thermodynamics. We hope that the new solutions described here may not only shed new light on the analytical structure of the nonlinear second-order hydrodynamic equations but also serve as a test of numerical  codes for the hydrodynamic simulation of heavy-ion collisions.

\textit{Acknowledgements}---The authors thank the Yukawa Institute for Theoretical Physics, Kyoto University, where this collaboration started during the YITP-T-13-05 workshop ``New Frontiers in QCD". We thank helpful conversations with the participants of this workshop. J.~N.\ thanks G.~S.~Denicol for enlightening discussions and Conselho Nacional de Desenvolvimento Cient\'ifico e Tecnol\'ogico (CNPq) and Funda\c c\~ao de Amparo \`a Pesquisa do
Estado de S\~ao Paulo (FAPESP) for financial support.


\begin{thebibliography}{99}
\bibitem{footnote1} $AdS_3$ is the space satisfying $X_0^2-X_1^2-X_2^2 + X_3^2 =L^2$. The Poincar\'e coordinates shown in (\ref{conformal}) are obtained by the parametrization $X_0=\frac{t}{x_\perp}L$, $X_1=\frac{z}{x_\perp}L$, $X_2= \frac{L^2 -r^2 +t^2}{2x_\perp}\,$, $X_3=\frac{L^2+r^2-t^2}{2x_\perp}$. The global coordinates are obtained by $X_0 = L\cosh \rho \cos \tau\,$, $X_1=L\sinh\rho \sin \Theta\,$, $X_2=L\sinh\rho \cos \Theta\,$, $X_3=L\cosh\rho \sin \tau$.
\end{thebibliography}
\end{document}